# AI prediction of cardiovascular events using opportunistic epicardial adipose tissue assessments from CT calcium score

Tao Hu[a], Joshua Freeze[a], Prerna Singh[a], Justin Kim[a], Yingnan Song[a], Hao Wu, PhD[a], Juhwan Lee, PhD[a], Sadeer Al-Kindi, MD, PhD[b,c], Sanjay Rajagopalan, MD, PhD[b,c], David L. Wilson, PhD[a,d], Ammar Hoori, PhD[a]

[a]Department of Biomedical Engineering, Case Western Reserve University, Cleveland, OH, 44106, USA
[b]Harrington Heart and Vascular Institute, University Hospitals Cleveland Medical Center, Cleveland, OH, 44106, USA
[c]School of Medicine, Case Western Reserve University, Cleveland, OH, 44106, USA
[d]Department of Radiology, Case Western Reserve University, Cleveland, OH, 44106, USA

**Abstract**

**Background:** Recent studies have used basic epicardial adipose tissue (EAT) assessments (e.g., volume and mean HU) to predict risk of atherosclerosis-related, major adverse cardiovascular events (MACE).

**Objectives:** Create novel, hand-crafted EAT features, "fat-omics", to capture the pathophysiology of EAT and improve MACE prediction.

**Methods:** We segmented EAT using a previously-validated deep learning method with optional manual correction. We extracted 148 radiomic features (morphological, spatial, and intensity) and used Cox elastic-net for feature reduction and prediction of MACE.

**Results:** Traditional fat features gave marginal prediction (EAT-volume/EAT-mean-HU/BMI gave C-index 0.53/0.55/0.57, respectively). Significant improvement was obtained with 15 fat-omics features (C-index=0.69, test set). High-risk features included volume-of-voxels-having-elevated-HU-[-50, -30-HU] and HU-negative-skewness, both of which assess high HU, which as been implicated in fat inflammation. Other high-risk features include kurtosis-of-EAT-thickness, reflecting the heterogeneity of thicknesses, and EAT-volume-in-the-top-25%-of-the-heart, emphasizing adipose near the proximal coronary arteries. Kaplan-Meyer plots of Cox-identified, high- and low-risk patients were well separated with the median of the fat-omics risk, while high-risk group having HR 2.4 times that of the low-risk group (P<0.001).

**Conclusion:** Preliminary findings indicate an opportunity to use more finely tuned, explainable assessments on EAT for improved cardiovascular risk prediction.

**Introduction**

Cardiovascular disease is a major cause of morbidity and mortality worldwide (1), leading to 17.9 million deaths globally each year (2). Numerous risk score methodologies have been developed to predict risks from cardiovascular disease, but these methods often lack sufficient discrimination (3). Accurate explainable risk prediction models will provide useful information to patients and physicians for more personalized medications and interventions.

Previous studies have determined the usefulness of coronary calcification Agatston score as obtained from CT calcium score (CTCS) images for cardiovascular risk prediction. The whole-heart Agatston calcium score has been found to be the best single predictor for major adverse cardiovascular events (MACE) (4–6), outperforming clinical factors such as smoking, blood pressure, and total cholesterol (7). However, in the early stages of coronary artery disease (CAD), calcifications maybe minimal or even non-existent. Moreover, CAD sometimes involves vulnerable plaque formation without a heavily calcified component to be captured in the Agatston score. One report indicates that half of patients under 40 years of age with obstructive CAD on coronary CT angiography (CCTA) have zero Agatston, highlighting the age-dependency of the prognostic value of Agatston (8). These findings suggest that there might be room for improvement.

The epicardial adipose tissue (EAT) has garnered significant attention for MACE prediction. Of particular interest is the pericoronary adipose tissue (PCAT), a subset of EAT located adjacent to the coronary arteries. Several studies indicated that PCAT volume and elevated fat attenuation index were associated with coronary inflammation and subsequent MACE (9–11). Quantitative features obtained from PCAT and EAT have been linked with cardiovascular risk but this has not been widely examined for prediction of MACE.

There is a pathophysiological rationale for the role of EAT in MACE risk. EAT is in vascular communication with the myocardium and coronary arteries (12,13). As a result, inflammation in EAT can affect atherosclerosis development due to the secretion of pro-inflammatory and profibrotic cytokines (14). It has been determined that inflammation of fat results in higher HU values (10,11,15,16). Notably, EAT is not uniformly distributed and has regional differences. The idea that EAT and PCAT can signal atherosclerosis development has been dubbed the 'outside-in theory' of atherosclerosis (17). Regarding EAT specifically, current studies have focused on simple features, including EAT volume, mean intensity level, and max thickness (18–24), potentially leaving room to identify other important characteristics of EAT. Table 1 reviews published reports of MACE and CAD prediction from EAT features.

In this preliminary investigation, we sought to use AI to predict MACE using features derived from EAT in non-contrast CTCS images. This includes automated deep-learning based EAT segmentation, extraction of novel features, Cox time-to-event modeling, analysis of high-risk features, and a combined EAT fat-omics model for MACE prediction.

**Methods**

**Dataset**

A total of 400 randomly selected patients without known coronary disease were identified from a clinical cohort undergoing clinical CTCS for risk assessment between 2014 and 2020. All CTCS scans were ECG-gated, non-contrast acquisitions, with standardized scanner settings of 120kVp and 30mAs. The slice thickness was 2.5mm, and the average in-plane pixel spacing was 0.5×0.5mm, with slice dimensions of 512 × 512 and an average of 50 slices per volume. Patients were followed for incidence of MACE (defined as myocardial infarction, stroke, coronary revascularization, and all-cause mortality) obtained from electronic health record review. The max follow-up time was 6 years follow up observation time. Times were measured from the time of imaging to event or censoring. The dataset in this preliminary study was MACE-enriched (56% of patients had MACE during follow-up) to improve precision of image-derived effects. Population characteristics are described in Table 2.

The dataset was randomly divided into training/held-out testing subsets (80%/20%), respectively, while maintaining equal MACE ratio in both subsets. Clinical data such as age, sex, body mass index (BMI), and Agatston score, were also collected. Some clinical data elements were missing (e.g., out of 400 patients, 339 had reported BMI).

### EAT segmentation

We emphasized accurate segmentation of cardiac fat depots. EAT is located inside the heart pericardium. Accurate manual segmentation is quite intensive, taking up to 2 hours per case with inter- and intra-reader variability. We leveraged our previously developed DeepFat (25), a deep learning-based automatic segmentation methods, which expedited the process and improved consistency. In this study, we ran DeepFat to get an initial segmentation and manually corrected any errors. The detailed pipeline of automatic segmentation is illustrated in central illustration figure A.

### Feature engineering

We created pathophysiologically-inspired, hand-crafted features, giving us a total of 148 features from EAT. Features were organized into three categories: morphological, intensity and spatial features (central illustration figure C). Examples of morphological features are volume, principal axis lengths and EAT thickness. Intensity-based features included statistical measurements, such as HU min, max, mean, skewness, and histogram bins. To analyze the spatial distribution of EAT within the heart region, we subdivided the heart region into four equally thick(axial) slabs of image slices from top to bottom (Fig. 1) and four equidistant ribbons from outside to inside (central illustration figure C).

To use features from the EAT thicknesses distribution, we developed a novel algorithm (Fig. 3A) that measured EAT thicknesses across the EAT 3D surface. Along each ray from the centroid of the heart to the pericardium surface, the algorithm measured the Euclidean distance between the first and last intersected fat voxels. Rays scan the whole 3D sac surface with a one-degree shift (i.e., 360-deg in the XY-plane and 180-deg in the Z-direction). This method provided 64,800 EAT thickness measurements for each patient. Subsequently, we computed statistical features from the distribution (e.g., mean, max, skewness). Additionally, we divided the thickness measurements into four fixed histogram bins (each 8mm wide), as determined by the spread of thickness observed across all persons. This allowed us to identify specific thickness ranges that may be particularly relevant for the prediction of MACE. Ultimately, our method provides a much more detailed and comprehensive assessment of EAT thickness as compared to previous studies.

We created features to analyze EAT elevated HU values, thought to be an indicator of adipose inflammation (10). This provided a more detailed analysis than the standard approach of computing mean HU (23,24), We used statistical measurements of the distribution of intensity values, including extremes (max, min), peakedness (via kurtosis) and asymmetry (via skewness) of the intensity distributions. In addition, we analyzed 4 and 8 histogram bins of HU values and extracted statistical metrics for each defined range.

We also created features associated with the spatial distribution of EAT. We divided the EAT volume into quartiles from bottom to top based on the number of axial slices. Similarly, we divided EAT volume into four quartile "shells" from outer to inner, based on the length from the major principal axis. This division was performed using morphological operations. This process divided the EAT volume into four quartiles sub-regions axially and four shells sub-regions radially. For each sub-region, we performed morphological and intensity-based analyses. Example sub-regions features are: EAT volume for the topmost 25% of the heart and intensity analyses for the outermost ribbon.

Compared to the state-of-the-art methods (18–24), our proposed novel feature extraction methods create a much more comprehensive evaluation of EAT.

**Feature selection and evaluation metrics.**

We performed feature reduction to limit correlated features and over-training. Starting with 148 EAT features, we applied maximum relevance minimum redundancy (MRMR) to remove correlated features, resulting in 50 fat-omics features. Next, we used Cox proportional hazard model with elastic regularization to select the most effective features (26). Specifically, we used elastic net regularization with a mixing parameter ($\alpha$) of 0.8, determined using grid search, to balance the strengths of the L1 (Lasso) and L2 (Ridge) penalties to achieve optimal performance. In addition, we used 10-fold cross-validation to determine the optimal hyperparameters and avoid overfitting. Our final model included 15 fat-omics features. All feature preselection and model manipulation were performed on the training set. The resulting model was then evaluated on both the training and held-out test set using various evaluation metrics, including C-index, Akaike's information criterion (AIC), hazard ratio and area under the ROC curve (AUC) at given time points. We also analyzed hazard ratios (HRs) and examined their significance using p-values. In addition, to identify explainable high-risk features and feature subsets, we designed additional univariate and multivariate Cox models for investigation and comparison purposes.

**Results**

We performed a comprehensive feature analysis to identify most influential features for predicting MACE within each feature category. We started with the spatial distribution of EAT. In Table 3, we compared the predictive ability of EAT volume and each slab sub-region using univariate cox models. Total EAT volume is predictive, but with a low C-index (C-index=0.53, p=0.005). The volume in the top slab of the heart is more predictive (C-index=0.56, p = 0.001). To further evaluate the significance between models, we added mean HU for both models (EAT volume + mean HU vs. EAT in the top slab (EAT_PQ4) + mean HU). Utilizing likelihood ratio test, the second model with EAT_PQ4 demonstrated a significantly better fit than the model with EAT volume, having higher log-likelihood (p<0.001). We performed similar analyses on shell sub-regions (Supplement Table S1). The volume in the outermost sub-region shows the most significance with a C-index of 0.56 with (p<0.001). These observations are consistent with analyses of PCAT which is usually analyzed in top heart region.

We also analyzed the role of HU values on MACE prediction. In Fig. 2A and 2D, we showed spatial distributions of HU values for example patients with and without a

MACE event, respectively. In the PCAT surrounding the RCA (Fig. 2B and 2E), the patient with MACE had clearly elevated HU values as compared to the patient without MACE, consistent with reports in the literature suggesting that elevated HU in PCAT is indicative of inflammation and increased risk of MACE (10,11). In the histograms (Fig. 2C and Fig.2F), the patient with MACE had a decided shift to high HU values relative to the patient without MACE, again emphasizing the effect of HU values. Using similar univariate Cox analyses, we further analyzed HU features across individuals (Table 4). The mean HU of EAT was a poor predictor of MACE. The volume of voxels between -50 and -30 HU, the highest bin of HU, was the most significant feature, emphasizing the importance of elevated HU values. Negative HU skewness was a significant predictor of MACE, again emphasizing the importance of high HU values. In addition to using the volumes of tissue within each range, we normalized histograms and obtained a probability of being within a range of HU values. Results (C-index = 0.55) were slightly worse compared to those obtained with absolute volumes (Supplement Table S2). Both normalized and unnormalized features were included in our starting comprehensive feature set.

We analyzed the role of EAT thicknesses on MACE prediction using univariate Cox models (Fig. 3). Figure 3A shows the process of measuring EAT thickness, where all thicknesses were determined along rays originating from the centroid of the pericardium to the 3D surface. We collected all measurements (Fig. 3B) and analyzed in terms of histograms (Fig. 3C). In the table, mean thickness was not a good predictor (C-index =0.49, p=0.44). Maximum thickness, histogram kurtosis, and histogram skewness were better. Notably, the largest thickness bin (24-32mm) emerged as the most predictive (C-index = 0.53, p=0.006) among all bins (Supplement Table S3). Together, investigations indicate that large thickness values were important for MACE prediction.

We analyzed the role of the shape of the heart as assessed from the pericardium segmentation on MACE prediction (Fig. 4). Figures 4A-D show the distribution of heart's principal axes and aspect ratio across MACE and no MACE groups. We found that the lengths of the major and intermediate axes were most predictive, with the minor axis length having little effect. This suggests that a large, flat sac volume is a risk factor.

In Table 5, we compared Cox models designed from combination of important fat-omics features to our proposed fat-omics model. We include results on both the training and held-out testing cohorts to enable determination of any. For baseline comparisons, we included predictions based upon (i.e., EAT_vol, mean_HU, and thickness_Max). These gave test C-index values not much above guessing (0.5). When we combined these three "traditional" features in line 4, prediction improved giving a test C-index of 0.6, suggesting independence of the features. Prediction from only 3 high-risk features previously discussed (Vol_PQ4, Pro_50_30, and Thickness_Kurtosis) further improved performance (test C-index=0.64). Finally, our full fat-omics model with the 15 fat-omics features identified in Supplement Table S4 gave by far the best results on this dataset. It gave the best performance values in each column, including test C-index and 2-year AUCs of 0.69 and 0.70, respectively.

In addition, we compared the ability of the combined fat-omics model to stratify high- and low-risk groups to that of the traditional EAT_vol Cox model (Fig. 5). In the testing cohort, the fat-omics model resulting in a significant reclassification of risk beyond EAT volume at year 2 (training: $NRI_{Categorical} = 0.259$ [95% $CI$ 0.10, 0.37; $p < 0.001$] ($NRI_{Non-event} = 0.153$ and $NRI_{Event} = 0.106$), and testing: $NRI_{Categorical} =$

$0.101 [95\% \ CI -0.17, \ 0.37; \ p = 0.463](NRI_{Non-event} = 0.044 \ and \ NRI_{Event} = 0.057$). The non-significant p-value in the testing cohort can be attributed to the small size of this cohort, which comprised only 80 patients. Detailed reclassification table was shown in Supplement Table S5.

To highlight the importance of fat-omics results, we showed two exampled cases (Fig. 6). Both cases have similar BMI and Agatston calcium scores of zero. The heart on the left has much higher EAT volume and vol_50_30 values than the one on the right. The fat-omics model predicted 3 times the risk for the heart on the left than the one on the right. There was a MACE event for the heart on the left but not the right.

**Discussion**

To the best of our knowledge, we have created the first-ever machine learning comprehensive analysis of EAT features (our fat-omics) for MACE prediction. Previous reports identify that EAT volume (22–24) and, EAT HU (23,24) are associated with MACE, and that the thickness of the right ventricular anterior free wall is correlated to obstructive CAD (21). We used features like these and investigated and our own hand-crafted features in analyses. Our comprehensive fat-omics risk prediction model with 15 hand-crafted features was much improved as compared to other models evaluated. For example, fat-omics gave testing C-index/2-year-AUC values of 0.69/0.70, while EAT volume and mean HU yielded only 0.53/0.50 and 0.49/0.55, respectively. High risk features in fat-omics included assessments of regional EAT distribution, EAT thickness assessments, pericardium shape, and high HU values, suggesting that EAT is indeed a risk factor for MACE. It appears that much more can be gleaned from a more detailed assessment of EAT than has been done in the past. This will lead to a path for important studies combining fat-omics features with coronary calcification assessments. Potentially, this will lead to much-improved risk prediction as compared to the Agatston score, the only current standard clinical assessment from CTCS images.

We identified multiple novel features important for MACE prediction. These include the EAT volume in the top quartile slab of the heart, the EAT volume of voxels within -50 to -30 HU, thickness kurtosis, and the major principal axis length. Importantly, the volume of voxels between -50 and -30 HU (vol_50_30) was the most MACE informative feature. Fat inflammation is linked to high HU, presumably due to morphological changes of adipocytes (27), inhibition of local adipogenesis (10,28), and increased in vascular permeability (27,29). See Fig. 2 for an example figure showing increased HU values with MACE. The increase in the highest of EAT HU values is elegantly captured with vol_50_30. The attribute of high HU values is also captured by negative skewness of the HU intensity distribution, again found to be a significant risk factor of MACE in time-to-event modeling.

There are some limitations to our preliminary study. Most importantly, we used a small MACE-enriched cohort. This was done so that we could emphasize feature engineering on a dataset with expert-proven, accurate EAT segmentations. By using a MACE-enriched cohort, we emphasize the importance of imaging features with a reduction of the uncertainty obtained with low event rate time-to-event modeling. Another limitation is that the MACE-free survival curves (Fig. 8) are dedicated for our MACE-enriched dataset only, and not for a general population. In addition, although we have identified explainable, very promising, novel features, there are always more features that could be created. For future investigation, we note that for EAT in CTCS images, there are no existing adequate radiomics-style libraries. Deep learning approaches may be able to capture more complex features and improve the predictive power of the model. However, the explainability of our results (e.g., the contribution of high HU adipose volume) encourages the hand-crafted feature approach.

In conclusion, our study highlights the potential of applying a detailed image-based analysis of EAT for improving the prediction of cardiovascular risk. We have demonstrated that this can be done opportunistically in lost-cost (no-cost) CT calcium score exams. It will be most interesting to combine AI analysis of adipose depots and coronary calcifications together for risk prediction. Given preliminary reports of the independence of these assessments (30),such an analysis holds promise. It will be interesting to determine if risk prediction from EAT is useful for identifying those patients with zero calcium who go on to have a MACE event. There are drugs like GLP1 agonists and SGLT2 inhibitors (31), which tend to reduce epicardial adipose tissue (32), suggesting a mechanism for cardio-protection that could be further analyzed with a fat-omics model. Future studies are needed to validate our findings and to explore the use of more complex machine learning approaches to improve the predictive power of the model.

**Declaration:** Human subject research has been done under an IRB of Case Western Reserve University (CWRU) and University Hospitals Health Systems (UHHS), Cleveland, OH. CT calcium score images were acquired at UHHS, de-identified, and shared under a data use agreement. Research was supported by the National Heart, Lung, and Blood Institute through grants R01HL167199, R01HL165218, and R44HL156811. The content of this report is solely the responsibility of the authors and does not necessarily represent the official views of NIH.

**Conflicts of interest.** The only potential conflicts of interest relevant to the technology described herein are pending CWRU and UH patents to analyze CT calcium score images. This information has been disclosed to CWRU, and PI DLW has an approved CWRU plan for managing any potential conflicts.

**Figure and table titles and legends**

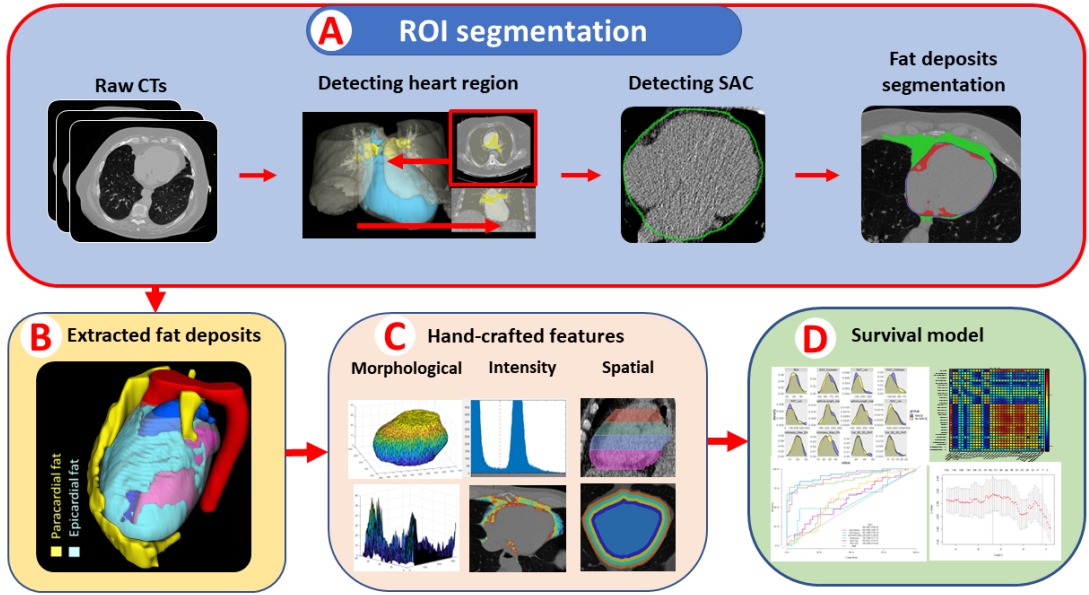

**Central illustration**: Pipeline for image analysis and MACE risk prediction. In (A), automatic semantic segmentation is performed to segment EAT, generating volumes of interest (VOIs) (B) for subsequent feature extraction. In (C), hand-crafted features, including intensity and spatial features, are extracted from VOIs. In (D), a variety of univariate and multivariate Cox proportional models are developed to assess the ability of individual features to predict MACE. Cox elastic net models with feature selection are employed to generate EAT-omics as an assessment for MACE risk prediction.

| Author | Modality | Cohort | Predictor | Outcome | Results |
|---|---|---|---|---|---|
| Cheng et. al 2010[1] | CT | 232 | EAT volume > 125 $cm^3$ | MACE | OR: 1.74[1.03-2.95] |
| Picard et. al 2014[2] | CT | 970 | EAT max thickness on LVLW>2.8mm | CAD | OR: 1.46 [1.03-2.08] |
| Goeller et. al 2018[3] | CT | 456 | EAT volume Mean HU | MACE | HR: 4.6[1.6-13.1] HR: 0.8[0.7-0.9] |
| Eisenberg et. al 2020[4] | CT | 2068 | EAT volume Mean HU | MACE | HR: 1.35[1.07-1.68] HR: 0.83[0.72-0.96] |
| Yang et al. 2023[5] | CT | 290 | EAT volume > 108.3 $cm^3$ | MACE | HR: 3.3[2.1-5.2] |
| Demircelik et. al 2014[6] | CCTA | 131 | EAT max thickness on RVAW | Obstructive CAD | AUC = 0.715 |
| Brandt et. al 2022[7] | CCTA | 117 | EAT volume | MACE | HR: 2.41 [1.08-4.72] c-index = 0.72 |
| Uygur et. al 2021[8] | CCTA | 127 | EAT volume EAT volume > 123.2 $cm^3$ | MACE for patients with type 2 diabetes | OR: 1.027[1.01-1.04] AUC = 0.82 |

*LVLW: Left ventricle Lateral Wall. RVAW: Right ventricle anterior wall

**Table 1**: Association of epicardial adipose tissue with MACE or coronary artery disease (CAD) in previous publications. Multiple studies have demonstrated that EAT volume,

mean HU, and thickness are predictive of MACE or CAD. Most entries should be self-explanatory. Previous studies have not included a detailed analysis of engineered features as we have done in this paper.

| Characteristic | Full cohort | MACE sub-cohort | No MACE sub-cohort | P-value |
|---|---|---|---|---|
| Patients | 400 | 224(56%) | 176(44%) | - |
| Female | 214(54%) | 112(50%) | 102(58%) | - |
| Age mean(min, max) | 61(18,87) | 63.2(18,87) | 58.3(21,84) | <0.001 |
| MACE events | 224(56%) | - | - | - |
| Observation time | 641.3 (10,2176) day 1.8 (0, 5.9) year | 505.3 (10,2170) day 1.4 (0,5.9) year | 814.5(14,2176) day 2.2 (0,5.9) year | <0.001 |
| Ag score | 602.3(0,5879) | 952.7(0,5897) | 156.3(0,3611) | <0.001 |
| Zero Ag score | 204(51%) | 116(52%) | 88(50%) | - |
| EAT vol (cm$^3$) | 120.7(20.5,399.4) | 127.4(26.8,399.4) | 112.2(20.5,300.1) | 0.004 |
| SAC vol (cm$^3$) | 774.9(391,1452.1) | 791.3(464.2,1312.1) | 754(391,1452.1) | 0.004 |
| EAT max thickness (mm) | 19.7(5.5,31.3) | 20.1(7.5,31.3) | 19.2(5.5,28.8) | 0.003 |

**Table 2**: Characteristics of the 400-patient cohort with an enriched MACE event rate (56%) and with proven EAT segmentation. Several assessments are significantly different between the groups.

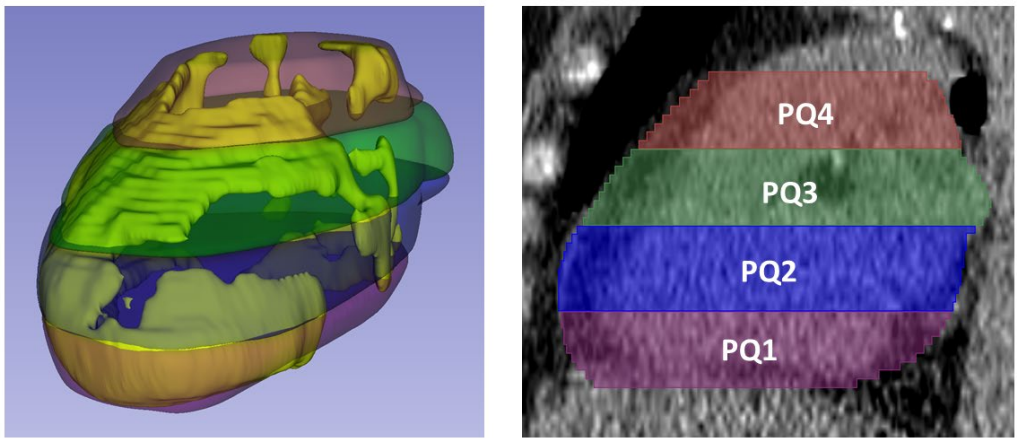

**Figure 1**: Slab volumes for localizing the spatial distribution of fat. The heart was divided into four equally thick slabs, each composed of consecutive axial image slices. Slabs extending from the lowermost to the uppermost parts of the heart, are labeled PQ1 to

PQ4. This slab-based approach allowed assessment of EAT features in different parts of the heart.

| Variables | C-index | AIC | HR[95%CI] | p-value |
|---|---|---|---|---|
| EAT_vol | 0.53 | 2345.81 | 1.2 (1.06:1.36) | 0.0046 |
| Vol_PQ1 | 0.53 | 2349.07 | 1.15 (1.01:1.3) | 0.037 |
| Vol_PQ2 | 0.53 | 2345.77 | 1.2 (1.06:1.35) | 0.0037 |
| Vol_PQ3 | 0.52 | 2349.05 | 1.15 (1.01:1.3) | 0.033 |
| **Vol_PQ4** | **0.56** | **2343.41** | **1.23 (1.09:1.39)** | **0.0012**\*\* |

**Table 3**: MACE prediction from EAT volume, including the contributions from different slabs. Each row presents a univariate Cox proportional hazards model for the respective feature, including the hazard ratio and corresponding p-value. Total EAT_vol is significant, but despite its prominence in the literature, its c-index is only 0.53. EAT in the top of the heart, corresponding to slab PQ4, is the most significant and gives a c-index of 0.56. Notably, this slab coincides with the region subjected to pericoronary adipose analysis.

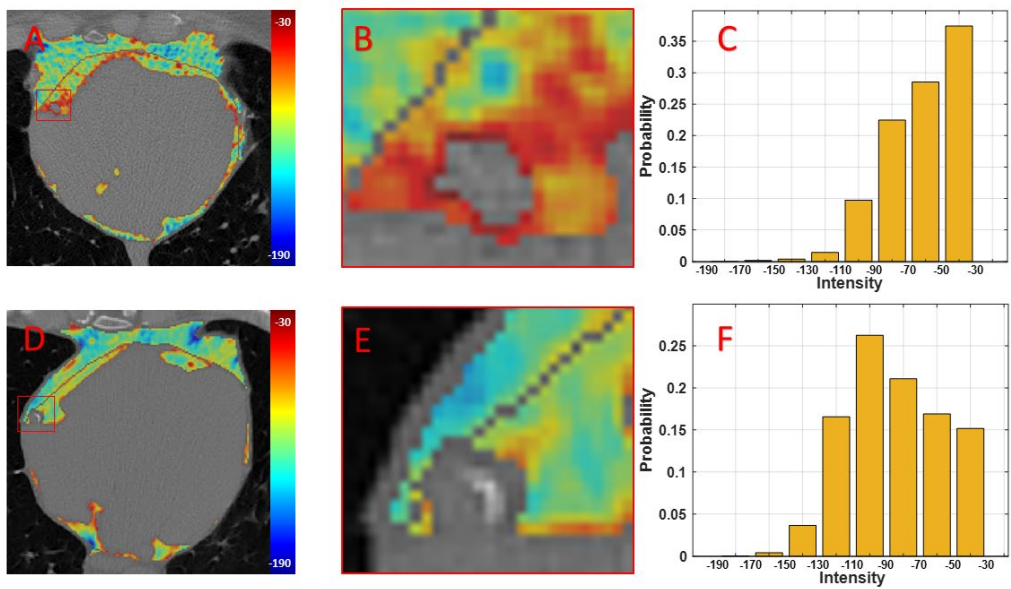

**Figure 2**: Example EAT CT intensities (HU values) in patients with and without MACE (top and bottom rows, respectively). In panels A and D, EAT HU (inside the pericardium

in red) is elevated as compared to PAT HU (outside the pericardium). Zoomed views show that the specialized EAT around the RCA (called PCAT) has elevated HU as compared to PCAT outside the black pericardium boundary. Importantly PCAT is clearly elevated in E relative to B. Panels C and F present normalized HU distributions for EAT. For the MACE patient (C), there is a conspicuous shift towards higher HU values with 37% of values within the [-50 HU, -30 HU] range, as compared to 15% for the no-MACE patient.

| Variables | C-index | AIC | HR[95%CI] | p-value |
|---|---|---|---|---|
| EAT_vol | 0.53 | 2345.81 | 1.2 (1.06:1.36) | 0.0046** |
| EAT_HUmean | 0.48 | 2353.26 | 1 (0.98:1.02) | 0.97 |
| vol_190_170 | 0.55 | 2352.59 | 1.05 (0.942:1.16) | 0.4 |
| vol_170_150 | 0.54 | 2351.59 | 1.07 (0.969:1.19) | 0.18 |
| vol_150_130 | 0.52 | 2351.07 | 1.09 (0.976:1.23) | 0.12 |
| vol_130_110 | 0.51 | 2349.98 | 1.12 (0.997:1.26) | 0.056 |
| vol_110_90 | 0.51 | 2349.10 | 1.15 (1.01:1.3) | 0.034* |
| vol_90_70 | 0.53 | 2347.57 | 1.18 (1.03:1.35) | 0.015* |
| vol_70_50 | 0.56 | 2343.63 | 1.23 (1.08:1.4) | 0.0015** |
| **vol_50_30** | **0.57** | **2338.47** | **1.29 (1.14:1.47)** | **<0.001*** |
| negative_skewness | 0.56 | 2350.00 | 1.13 (1:1.27) | 0.049* |

**Table 4.** Analysis of EAT intensity by assessing volumes within HU ranges. In the first 8 rows, HU ranges are analyzed where Vol_190_170 corresponds to the volume of EAT having HU values between -190 and -170 HU. HR values increase with higher HU values. The c-index, AIC, and p-values indicate improved prediction at higher HU bins, with the highest HU bin, Vol_50_30, being the most significant. The distribution metric, negative_skewness allows one to capture the tendency to high HU values and is also significant. The mean HU value, EAT_HUmean, is a poor predictor.

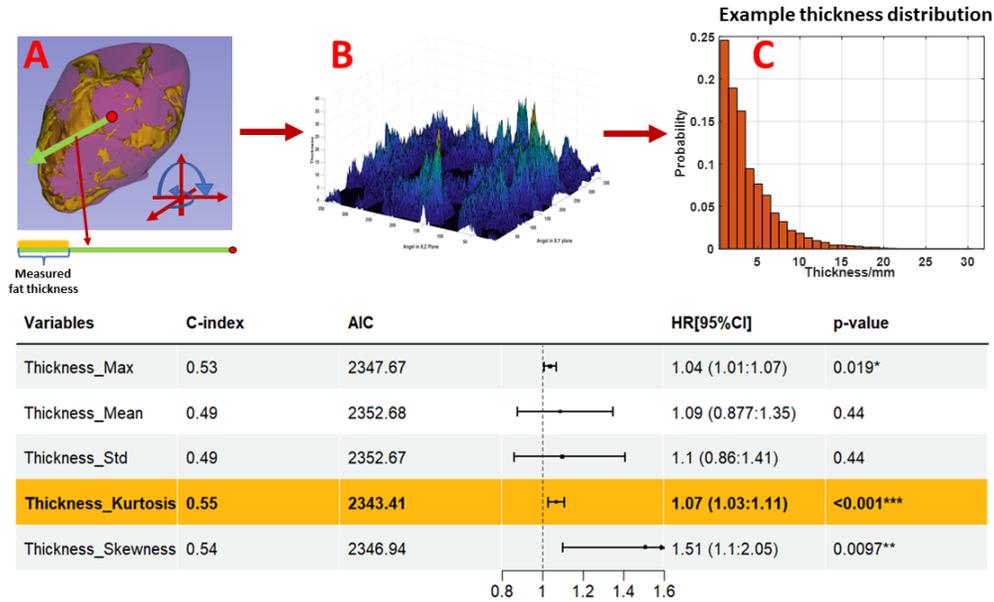

**Figure 3.** Analysis of the role of EAT thicknesses. Euclidian "thickness" was measured along a ray emanating from the center of mass of the pericardium volume (A). The ray was swept along all possible angles at one-degree increments to create a 2D array of thickness values (B), which were then gathered to create a histogram (C). On the bottom, we present univariate Cox proportional hazard models for features derived from thickness histograms. Mean thickness was not a good predictor. Kurtosis, indicating the spread of a histogram, and skewness, indicating the asymmetry of a histogram, both were much more predictive. The positive HR value for kurtosis emphasizes the importance of long-tail outliers.

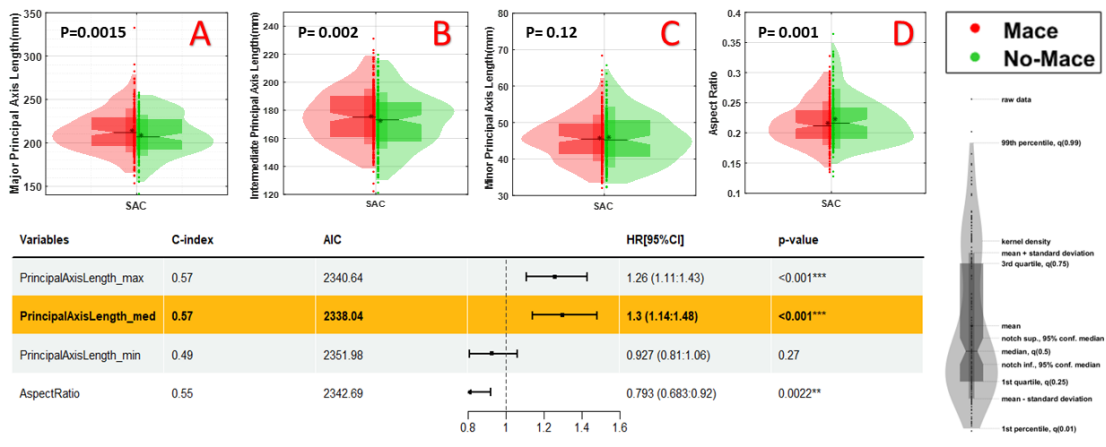

**Figure 4.** Shape features as assessed by principal axes of the pericardium volume. Along the top (A-C), we show violin plots of the major, intermediate, and minor axes for MACE

(red) and no-MACE (blue) groups, where p values were obtained from a two-sample t-test. Aspect ratio is shown in (D). The major (A) and intermediate (B) axes were significantly different for MACE and no-MACE. These findings are corroborated by the univariate Cox analysis in the table at the bottom.

| Model specification | | Training (N = 320) | | | | | Testing (N = 80) | |
| --- | --- | --- | --- | --- | --- | --- | --- | --- |
| Model | Feature | HR[95%CI] | C-index | P-value | AUC 2-year | AIC | C-index | AUC 2-year |
| EAT vol | ln(EAT_vol) | 1.2(1.0-1.4) | 0.53 | 0.026* | 0.53 | 1895.40 | 0.53 | 0.50 |
| EAT mean HU | EAT_mean_HU | 1(0.98-1.02) | 0.49 | 0.99 | 0.48 | 1901.07 | 0.55 | 0.57 |
| EAT thickness | Thickness_Max | 1.1(1-1.1) | 0.54 | 0.007** | 0.56 | 1893.84 | 0.51 | 0.52 |
| EAT traditional features | ln(EAT_vol)+ EAT_mean_HU+ Thickness_Max | 1.6(0.86-2.9) 1.0(1-1.1) 1.1(1-1.1) | 0.57 | 0.142 0.021* 0.069 | 0.60 | 1891.21 | 0.6 | 0.58 |
| EAT high risk features | ln(Vol_PQ4) + Pro_50_30 + Thickness_Kurtosis | 2(1.3-1.7) 28(1.3-588) 1.04(1.01-1.09) | 0.6 | <0.001*** 0.032* 0.044* | 0.66 | 1883.44 | 0.64 | 0.69 |
| **EAT-omics** | **EAT-omics(15 features)** | 2.7(2.1-3.5) | **0.66** | <0.001 *** | **0.72** | **1864.66** | **0.69** | **0.70** |

**Table 5**: Risk prediction from our aggregated EAT-omics model as compared to predictions from subsets of features, including "traditional" ones. In lines 1-3, we evaluated EAT volume, mean HU, and maximum thickness. Despite their appearance in previous publications, the predictive capability of these features was marginal. The combination of these features improved performance (line 4). A model with 3 previously identified high-risk features improved prediction (line 5). Each of these high-risk features had a positive effect HR and significant p values, highlighting their importance. The EAT-omics model surpassed all other assessed models with the highest c-index, 2-year AUC, and lowest AIC for both training and held-out testing set.

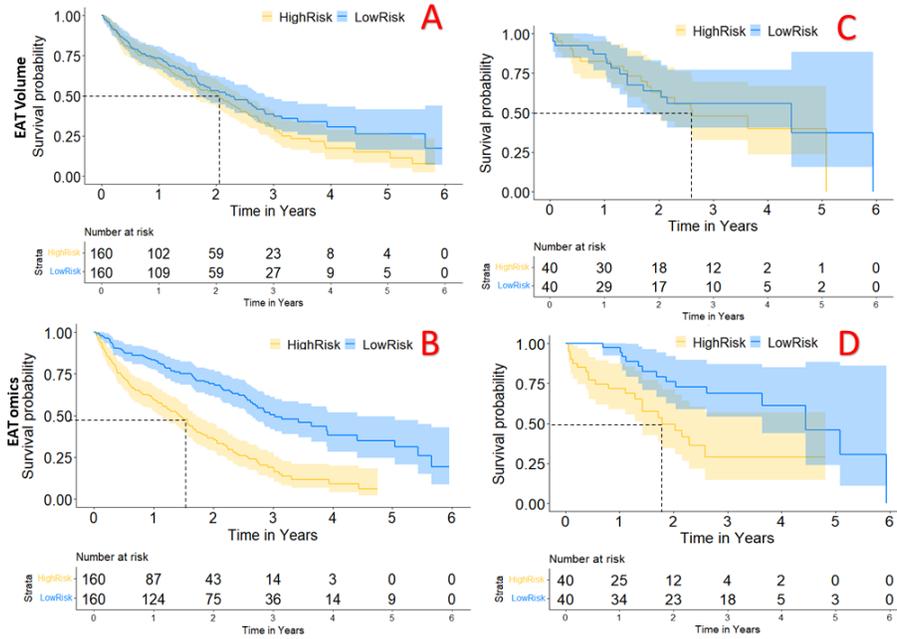

**Figure 5.** MACE risk stratification for our MACE-enriched dataset. We analyzed both the EAT_vol (top row) and EAT-omics (bottom row) models for the full cohort including training data (left column) and the much smaller testing cohort (right column). We stratified high- and low-risk cases on the median risk and then computed Kaplan-Meier. Mean test survival times for the high-risk groups are 1.78 years and 2.62 years for the EAT-omics and EAT_vol, respectively. The EAT-omics outperformed EAT_vol in both training and testing sets in terms of NRI. Please note that we use a small MACE-enriched cohort to engineer features. These small probabilities of MACE-free survival will not hold for the general population.

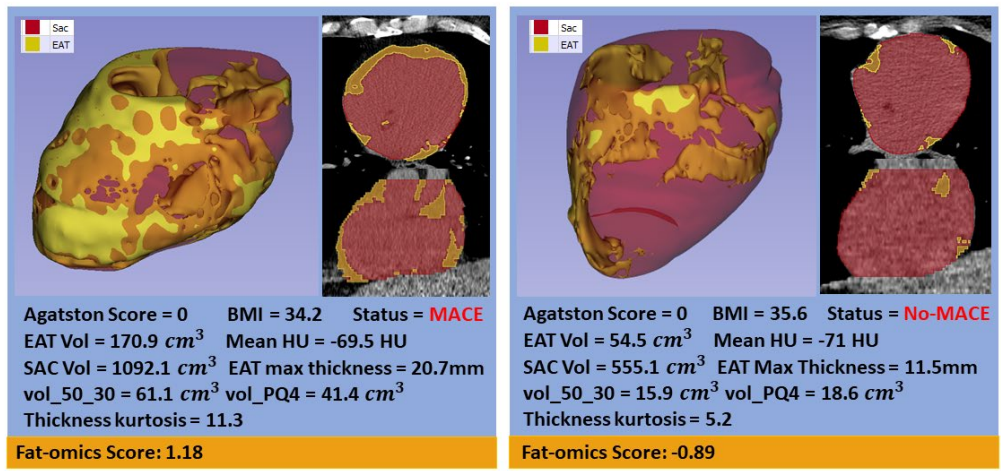

**Figure 6.** Comparison of cases having similar BMIs and Agatston of zero but having much different EAT profiles. On the left, the heart has higher EAT, sac, mean HU, and maximum EAT thickness than for the "lean" heart on the right. Our identified high-risk features (vol_50_30, vol_PQ4, and thickness kurtosis) shows even more difference. Applying the EAT-omics model, the risk for the left heart was 3-times that for the one on the right. In the case of the left heart, there was a MACE event later in the study.

**Supplemental Materials:**
**S.1 Detailed feature engineering**

Fat-omics involves precise measurements of heart structure and epicardial adipose tissue (EAT). The heart is equally divided into four axial slabs (Positional Quartiles (PQ)) from the top (PQ1) to the bottom (PQ4). EAT Hounsfield unit quartiles (HQ), ranging from 1 to 4, categorize HU values into bins: HQ1 includes values from -190 to -150, HQ2 from -150 to -110, HQ3 from -110 to -70, and HQ4 from -70 to -30. Spherical regions (SR) consisted of equidistant radial shells from the outside (SR1) to the inside (SR4) of the heart. We divided the thickness measurements into four fixed histogram bins (each 8mm wide).

1. *Structural features*
   - **Total_SAC_Volume_Cm3** (total pericardial sac volume in $cm^3$)
   - **WHV** (whole heart volume: pericardial sac volume – EAT volume)
   - **PrincipalAxisLength_max** (major principal axis length of the pericardial sac)
   - **PrincipalAxisLength_min** (minor principal axis length of the pericardial sac)
   - **PrincipalAxisLength_med** (intermediate principal axis length of the pericardial sac)

2. *EAT features*
   - **Total_Volume_Cm3** (EAT volume in $cm^3$)
   - **Total_Normalized_EAT** (EAT volume/pericardial sac volume)
   - **Total_HU<<stats>>** (<<mean, median, max, min, kurtosis, skewness>> of the HU of EAT)
   - **Thickness_<<stats>>** (<<mean, median, max, min, std>> of the EAT thickness)
   - **NormThickness_<<stats>>** (<<mean, median, max, min, std>> of the EAT thickness divided by the corresponding radius in the same direction)
   - **Thickness_bin<<number>>_Pro** (probability of fixed thickness bins)
   - **<<stats>>HU_PQ<<number>>** (<<mean, median, max, min, std, kurtosis, skewness>> of the HU in each positional quartile <<number>> [1-4])
   - **Vol_PQ<<number>>** (volume of EAT in $cm^3$ in each position quartile <<number>> [1-4])
   - **<<stats>>HU_HQ<<number>>** (<<mean, median, max, min, std, kurtosis, skewness>> of EAT HU in each HU bin HQ <<number>> [1-4])
   - **PixelCount_HQ<<number>>** (number of voxels of EAT in HQ <<number>> [1-4])
   - **Probability_HQ<<number>>** (probability of EAT Voxels with HU in HQ <<number>> [1-4])
   - **Vol_HQ<<number>>** (EAT volume in each HQ <<number>> [1-4])
   - **Pro_HQ<<HU range>>** (probability of EAT voxels in each HQ <<number>> [1-4])
   - **SR<<number>>_Pro_<<HU range>>** (probability of EAT voxels in HU ranges <<HU range>> ([-190, -170], [-170, -150], [-150, -130], [-130, -110], [-110, -90], [-90, -70], [-70, -50], [-50, -30]) in each spherical region <<number>>[1-4])

## S.2 Radiomics feature analyses

We provided additional detailed analyses into fat-omics features that correlate with the risk of major adverse cardiovascular events (MACE). These analyses included a range of EAT features, stratified by spherical regions (SR), Hounsfield Unit (HU) bins, and fixed thickness bins, to uncover the most significant predictors of MACE.

Figure S1 expanded on the predictive power of EAT volume from different spherical regions (SR), with special emphasis on the outermost layer (SR1), which demonstrated the highest significance in MACE prediction. This aligned with the anatomical fact that SR1 are proximal to the coronary arteries, also subjecting to pericoronary adipose analysis. Table S2 delved into the EAT HU normalized distributions, examining the probability of voxel volumes within specific HU ranges. The highest bin (Pro_50_30) still showed the highest significance among all bins. Table S3 presented analysis of EAT fixed thickness bins, offering insights into the distribution of thickness across predefined bins and their relationship to MACE, thus revealing the importance of EAT thickness heterogeneity in cardiovascular risk assessment. Table S4 consisted of all selected EAT features by cox elastic net, underscoring the intricate relationship between EAT characteristics and cardiovascular health. All high-risk features mentioned in the main text were included in final fat-omics model.

| Variables | c-index | AIC | HR[95%CI] | p_value |
|---|---|---|---|---|
| EAT_vol | 0.53 | 2345.81 | 1.2 (1.06:1.36) | 0.0046** |
| **Vol_SR1** | **0.56** | **2341.32** | **1.26 (1.11:1.43)** | **<0.001*** |
| Vol_SR2 | 0.54 | 2345.04 | 1.21 (1.07:1.37) | 0.0029** |
| Vol_SR3 | 0.53 | 2347.57 | 1.17 (1.03:1.32) | 0.013* |
| Vol_SR4 | 0.53 | 2351.07 | 1.1 (0.975:1.24) | 0.12 |

**Table S1:** MACE prediction from EAT volume, including the contributions from different shells sub-regions. Each row presents a univariate Cox proportional hazards model for the respective feature, including the hazard ratio, AIC, and corresponding p-value. The total EAT volume was divided into four quartiles shells from outer to inner (SR1-SR4), where the Vol_SR1 represents to EAT volume at the most outer shell. Similar as slabs sub-regions analyses in Table 3, EAT in the most outer shell of the heart is the most significant with a C-index of 0.56. This aligns with the anatomical fact that SR1 are proximal to the coronary arteries, also subjecting to pericoronary adipose analysis.

| Variables | c-index | AIC | HR[95%CI] | p_value |
|---|---|---|---|---|
| Pro_190_170 | 0.53 | 2351.99 | 1.07 (0.955:1.2) | 0.24 |
| Pro_170_150 | 0.52 | 2351.27 | 1.09 (0.968:1.24) | 0.15 |
| Pro_150_130 | 0.5 | 2352.30 | 1.07 (0.939:1.21) | 0.32 |
| Pro_130_110 | 0.48 | 2353.25 | 1.01 (0.885:1.15) | 0.9 |
| Pro_110_90 | 0.54 | 2350.23 | 0.886 (0.772:1.02) | 0.083 |
| Pro_90_70 | 0.54 | 2344.82 | 0.822 (0.719:0.938) | 0.0037** |
| Pro_70_50 | 0.48 | 2353.26 | 0.995 (0.872:1.13) | 0.94 |
| Pro_50_30 | 0.55 | 2350.26 | 1.13(1.001:1.29) | 0.05* |

**Table S2:** Analysis of EAT HU normalized distribution. In additional to just using the volumes of tissue within each range, we normalized histograms and obtained a probability of being within a range of HU values. Probability of EAT corresponding to HU ranges are analyzed where Pro_190_170 corresponds to the probability of EAT having HU values within -190 to -170 HU. Results were only a little poorer than obtained in Table 4, while the highest bin still showed significance.

| Variables | c-index | AIC | HR[95%CI] | p_value |
|---|---|---|---|---|
| thickness_bin41_Pro | 0.49 | 2353.03 | 0.968 (0.851:1.1) | 0.63 |
| thickness_bin42_Pro | 0.48 | 2353.18 | 1.02 (0.895:1.16) | 0.77 |
| thickness_bin43_Pro | 0.52 | 2350.73 | 1.1 (0.988:1.22) | 0.083 |
| thickness_bin44_Pro | 0.53 | 2347.79 | 1.19 (1.05:1.35) | 0.0059** |

**Table S3**: Fixed histogram bin analysis of EAT thickness. Similar as HU, we delved deeper by dividing the thickness measurements into four fixed histogram bins (each 8mm wide), determined by the spread of thickness observed across all patients. Probability of thickness are analyzed where thickness_bin44_Pro indicating the probability of thickness between 24 to 32mm. As expected, the largest thickness bin is the most significant feature, consistent with our findings in Fig. 4 (long tail thickness outlier is more significant).

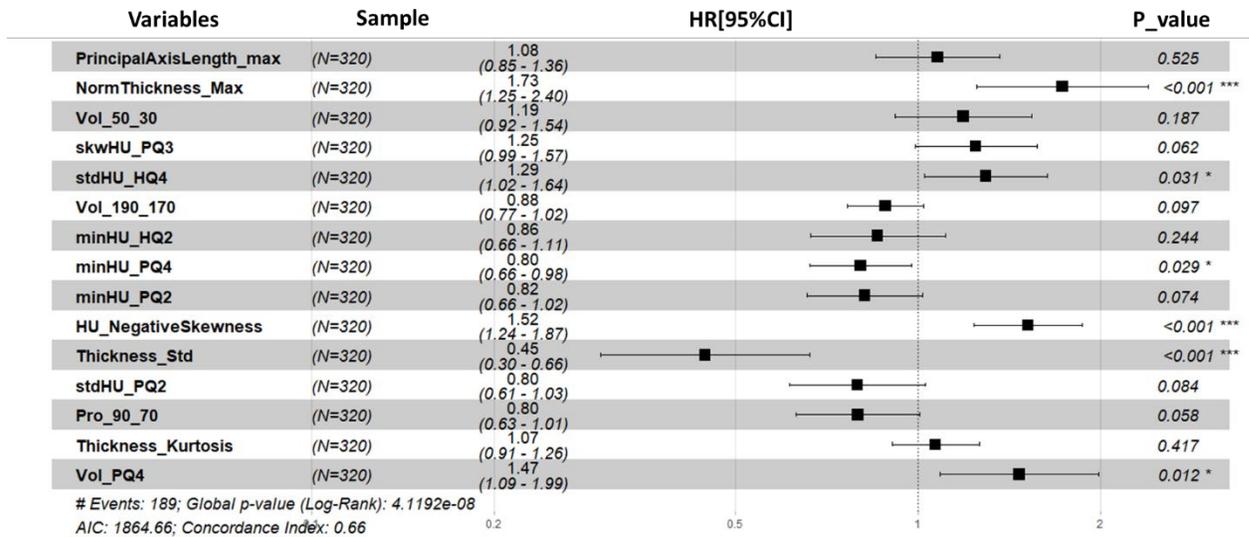

Table S4: List of all selected EAT features in EAT-Omics.
We started with 148 EAT features and performed maximum relevance minimum redundancy (MRMR) to exclude highly correlated features, resulting 50 uncorrelated EAT features. Among all preselected features, 15 dominant features were selected by cox elastic net. We performed cox proportional hazard analysis on those 15 selected features. Corresponding hazard ratio and p-value were listed in the table. All significant features introduced previously were included in this table.